\documentclass[12pt]{article}
\usepackage[T1]{fontenc}
\usepackage{bm}
\usepackage{cite}
\usepackage{mathrsfs}
\usepackage{slashed}
\usepackage{graphicx} 
\usepackage{hyperref}
\usepackage{verbatim}
\usepackage{color}
\usepackage{enumerate}
\usepackage{authblk}
\usepackage[hang,flushmargin]{footmisc}
\usepackage{lipsum}
\usepackage[font=footnotesize]{caption}
\usepackage[utf8]{} 
\usepackage{empheq}
\usepackage{amssymb,amsmath,amsfonts}
\usepackage[makeroom]{cancel}
\usepackage{hyperref}
\usepackage{color}
\newcommand{\cev}[1]{\reflectbox{\ensuremath{\vec{\reflectbox{\ensuremath{#1}}}}}}

﻿
﻿
\numberwithin{equation}{section}

﻿
\definecolor{blue-violet}{rgb}{0.54, 0.17, 0.89}
\definecolor{PineGreen}{cmyk}{0.92, 0, 0.59, 0.25}
\definecolor{OliveGreen}{cmyk}{0.64, 0, 0.95, 0.40}
\definecolor{RawSienna}{cmyk}{0, 0.72, 1, 0.45}
\definecolor{Gray}{cmyk}{0, 0, 0, 0.50}
\definecolor{MidnightBlue}{cmyk}{0.98, 0.13, 0, 0.43}
\definecolor{Orange}{cmyk}{0, 0.61, 0.87, 0}
\definecolor{LimeGreen}{cmyk}{0.50, 0, 1, 0}
\definecolor{Green}{cmyk}{1, 0, 1, 0}

﻿
﻿

﻿

﻿

﻿

﻿
﻿
\renewcommand{\tilde}{\widetilde}

﻿

\usepackage{mathrsfs}

﻿
﻿
﻿
\begin{document}
﻿
\title{\!\!\LARGE{Integrable black hole dynamics in the asymptotic structure of AdS$_3$}\vspace{10pt}}
﻿
﻿
\author[a]{\normalsize{Marcela C\'ardenas}\footnote{\href{mailto:marcela.cardenasl@uss.cl}{marcela.cardenasl@uss.cl}}}
\author[b]{\normalsize{Francisco Correa}\footnote{\href{mailto:francisco.correa.s@usach.cl}{francisco.correa.s@usach.cl}}}
\author[b,c]{\normalsize{Miguel Pino}\footnote{\href{mailto:miguel.pino.r@usach.cl}{miguel.pino.r@usach.cl}}}
﻿
\affil[a]{\footnotesize\textit{Universidad San Sebasti\'an, Facultad de Ingenier\'ia,  Bellavista 7, Recoleta, Santiago, Chile.}}
\affil[b]{\footnotesize\textit{Departamento de F\'{i}sica, Universidad de Santiago de Chile, Avenida Victor Jara 3493, Estaci\'on Central, 9170124, Santiago, Chile.}}
\affil[c]{\footnotesize\textit{Center for Interdisciplinary Research in Astrophysics and Space Exploration (CIRAS), Universidad de Santiago de Chile, Avenida Libertador Bernardo O'Higgins 3363, Estación Central, Chile.}}
\date{}
﻿
\maketitle
﻿
\begin{abstract}
This work deepens the study of integrable asymptotic symmetries for AdS$_3$. They are given by an infinite set of integrable nonlinear equations known as the Ablowitz-Kaup-Newell-Segur (AKNS) hierarchy, characterized by an also infinite set of abelian conserved charges. We present their field-dependent Killing vectors and the computation of the canonical charges associated to the asymptotic metric, together with their corresponding charge algebra. We study black hole thermodynamics and show that the temperature for stationary black holes falling in the AKNS asymptotics is always constant, even in the case where the solutions are not axisymmetric. This is related to the existence of a hyperelliptic curve, which appears as a fundamental object in many integrable systems. We also present a special solution associated with the Korteweg-de Vries equation, that is a particular case of the AKNS integrable hierarchy. It is presented in the form of a periodic soliton leading to a cnoidal KdV black hole, whose temperature is characterized by two copies of hyperelliptic curves.
\end{abstract}
﻿
\newpage
﻿
﻿
\section{Introduction} 
﻿
﻿
Since the foundational work of Brown and Henneaux \cite{Brown:1986nw},  several extensions and enhancements have been explored for the asymptotic structure of three-dimensional gravity, highlighting how non-trivial boundary conditions compensate for the lack of bulk dynamics.  Moreover, the simplicity of the Chern-Simons formulation of gravity\cite{cs1,cs2} have promoted the further understanding of its boundary degrees of freedom, where the topic of constructing asymptotic symmetries have been studied with extensive interest and from various viewpoints \cite{Compere:2013bya,Troessaert:2013fma,Avery:2013dja,Compere:2015knw,Apolo:2014tua,Grumiller:2017jft,Henneaux:2019sjx,Adami:2020ugu,Geiller:2021vpg}. 
Their applications rely on the holographic conjecture, proving the duality from black hole physics \cite{Banados:1998gg,Carlip,Banados1998-2,Bunster2014,Afshar:2016wfy,Afshar:2016kjj,Grumiller:2019tyl} to condensed matter \cite{Zee:1995avy,Wen:1995qn,suskind,Polychronakos,Witten:2015aoa}.
﻿In this article, we focus on a relaxed set of AdS$_{3}$ boundary conditions for the gravitational field,
\begin{subequations}\label{asym1}
  \begin{align}
     g_{tt}	&= g_{tt}^{(2)}(t,\phi)\rho^{2}+g_{tt}^{(0)}(t,\phi)+\mathcal{O}(\rho^{-2}),\\
g_{t\rho}&	=g_{t\rho}^{(-1)}(t,\phi)\rho^{-1}+\mathcal{O}\left(\rho^{-4}\right),\\
g_{t\phi}&	= g_{t\phi}^{(2)}(t,\phi)\rho^{2}+g_{t\phi}^{(0)}(t,\phi)+\mathcal{O}(\rho^{-2}),\\
g_{\rho\rho}&=\frac{\ell^{2}}{\rho^{2}}+\mathcal{O}\left(\rho^{-4}\right),\\
g_{\rho\phi}&	=g_{\rho\phi}^{(-1)}(t,\phi)\rho^{-1}+\mathcal{O}\left(\rho^{-4}\right),\\
g_{\phi\phi}&=g_{\phi\phi}^{(2)}\rho^{2}+g_{\phi\phi}^{(0)}+\mathcal{O}\left(\rho^{-2}\right),
\end{align}
\end{subequations}
where $\rho$ represents a radial coordinate, the temporal and angular coordinates are $t$ and $\phi$, respectively, while $\ell$ stands for the AdS radius.  Originally, \eqref{asym1} was derived by allowing, in full generality, an arbitrary number of fields in the Chern-Simons formulation of the theory \cite{Grumiller:2016pqb},  using the radial gauge \cite{Coussaert:1995zp}.  This behaviour additionally allows the leading order in the $g_{\phi\phi}$ component to vary,  as proposed in \cite{Troessaert:2013fma},  that is associated to a relaxation on the conformal boundary degrees of freedom.  More recently, it has also been shown to have consequences in the holographic stress tensor leading to Weyl currents \cite{Ciambelli:2019bzz,Alessio:2020ioh,Ciambelli:2023ott,Jia:2021hgy,Arenas-Henriquez:2024ypo}, encoded in the radial relaxation $g_{\rho a}$ for
$x^{a}=(t,\phi)$, namely Weyl-Fefferman-Graham gauges. 
The aim of this article is to show how to detour from the previous choices to realize an equally exciting but radically different asymptotic scenario.  In practice,  this is possible due to the fact that not all of the coefficients in \eqref{asym1} correspond to dynamical functions, as some are associated with Lagrange multipliers whose boundary behavior will be chosen as state-dependent, i.e., local functions of the dynamical fields. This scenario has been exploited to search for the mentioned non-trivial field-dependent symmetries of the Weyl charges,  nonetheless here,  they are constructed to connect gravity with an infinite family of 1+1 integrable structures.  Indeed, the Gardner hierarchy, Korteweg-de Vries (KdV), modified Korteweg-de Vries (mKdV), or Harry-Dym \cite{pttt, Afshar:2016kjj, Afshar:2016wfy, Erices:2019onl ,Grumiller:2019tyl, Dymarsky:2020tjh,opt,  Lara:2024cie, Cardenas:2024hah} are some of the possible allowed choices.  In what follows, we deepen in \cite{Cardenas:2021vwo}, where the connection with integrable hierarchies was proven with robust generality and also provide further insights on black hole thermodynamics.  
﻿\newline
Integrable systems and their associated techniques have played an interesting role in gravitational physics, for instance, through the construction of gravitational solitons using the Belinski–Zakharov method and the Ernst equation \cite{gravsol}, or in the analysis of isospectral properties in the perturbations of Schwarzschild black holes (see, for example, \cite{Glampedakis}). In this work, however, we shift our focus to the connection between integrable models and asymptotic symmetries, which provides a richer framework to uncover the underlying integrable structure. Many nonlinear integrable differential equations exhibit remarkable features, including the following:
\begin{itemize}
\item The inverse scattering method allows solving the initial value problem using linear equations related to quantum models.
\item There are different types of soliton solutions, including localized, periodic, topological, and combinations of these features.
\item Backlund and Darboux transformations act as field transformations mapping different pseudo-spherical surfaces and soliton solutions.
\item They have an infinite number of conserved charges.
\end{itemize}
In the forthcoming sections, we will show how these features emerge in various contexts within the asymptotic structure of AdS$_3$. In particular, an infinite family of integrable hierarchies appears, including the most well-known ones. Their corresponding solutions not only exhibit features associated with quantum linear problems, but also reveal a direct link between asymptotic symmetries and the infinite tower of conserved charges intrinsic to integrability. We follow \cite{cclp} where, by using the well-known relation between Chern-Simons and three-dimensional gravity \cite{cs1,cs2}, it was shown that the connection between AdS$_3$ gravity and integrable models includes the famous Ablowitz-Kaup-Newell-Segur (AKNS) system \cite{Ablowitz:1973zz, bookakns}, a mother theory for several of the most famous nonlinear integrable equations. Our main goal is to explore such connection further, using the metric formulation to highlight results hindered by the Chern-Simons formulation. It is demonstrated how the properties of the integrable system characterize the temperature of black hole solutions and explore its relation to quantum models.
﻿\newline
This paper is organized as follows: In Section 2, it is shown how the AKNS integrable structure arises from the asymptotic behavior \eqref{asym1} and review the most important features of the AKNS integrable nonlinear hierarchy.  Section 3 is devoted to the construction of the asymptotic symmetries and canonical generators of \eqref{asym1}. We focus in the AKNS boundary conditions,  explore the symmetries of the AKNS geometry and evince how they provide an intrinsic manner to functionally integrate the canonical charges.  Finally,  in section 5, we study the solution spectrum and show the existence of novel black hole solutions with non-trivial conserved charges,  associated to a cnoidal wave KdV dynamics.  We also discuss the thermodynamics of the AKNS black holes, where it was proven how they share some features with the quantum associated linear problem.
﻿
﻿
﻿
\section{Geometrization of AKNS}
The relationship between geometry, surfaces, and integrable systems has a long and beautiful history, starting with the key discovery of the sine-Gordon equation in the nineteenth century through the study of pseudospherical surfaces \cite{bookdar}. As shown in \cite{SASAKI1979343}, a similar construction can be performed for the AKNS hierarchy using the Gauss-Codazzi equations of a family of two-dimensional surfaces with constant negative Gaussian curvature. The approach adopted in the present work has the same spirit with a different realization: the integrable dynamics is introduced by the asymptotic behavior of a three-dimensional Einstein manifold.
﻿
In the following, we present a family of three-dimensional geometries satisfying the boundary conditions \eqref{asym1}, where the dynamics imposed by Einstein's equation corresponds to the AKNS system. This AKNS metric belongs to the relaxed $AdS_{3}$ boundary conditions \eqref{asym1}, with a precise choice of the metric coefficients. 

\begin{subequations}\label{metricaAKNS}
\begin{align} 
g_{tt} & =-\frac{B^{+}B^{-}\rho^{2}}{\ell^{2}}+\left(A^{+}{-}A^{-}\right)^{2}{+}B^{+}C^{+}{+}B^{-}C^{-}-\frac{\ell^{2}C^{+}C^{-}}{\rho^{2}} \quad,\quad\\ 
g_{t\rho} &=\frac{\ell(A^{-}{-}A^{+})}{\rho}\ ,\\ \notag
g_{t\phi} & =\frac{(B^{-}p^{+}{-}B^{+}p^{-})\rho^{2}}{2\ell}+\frac{1}{2}\ell\left[2(\lambda^{-}{+}\lambda^{+})(A^{+}{-}A^{-}){-}B^{+}r^{+}{+}B^{-}r^{-}{-}C^{+}p^{+}{+}C^{-}p^{-}\right]\\
 & +\frac{\ell^{3}(C^{-}r^{+}{-}C^{+}r^{-})}{2\rho^{2}} \  ,\\
g_{\rho\rho} & =\frac{\ell^{2}}{\rho^{2}}, \\
g_{\rho\phi} &=-\frac{\ell^{2}(\lambda^{+}{+}\lambda^{-})}{\rho}\  ,\\
g_{\phi\phi} & =p^{+}p^{-}\rho^{2}+\ell^{2}\left[p^{+}r^{-}{+}p^{-}r^{+}{+}(\lambda^{+}{+}\lambda^{-})^{2}\right]+\frac{\ell^{4}r^{+}r^{-}}{\rho^{2}} \ .
\end{align}
\end{subequations}
All functions $p^{\pm},r^{\pm}$ and $A^{\pm},B^{\pm},C^{\pm}$ depend on $t$ and $\phi$, except for $\lambda^{\pm}$, which are constants without variation and, as we will see next, are known on the integrable side as spectral parameters. 
This metric \eqref{metricaAKNS} can be expressed in the following form:
\begin{eqnarray}\label{compac}
ds^2&=&\left(\frac{\rho}{\ell}[ p^+\ell d\phi-B^+dt ]+\frac{\ell}{\rho}[ r^-\ell d\phi+C^- dt]\right)\left(\frac{\rho}{\ell}[ p^-\ell d\phi+B^-dt ]+\frac{\ell}{\rho}[ r^+\ell d\phi-C^+ dt]\right) \notag \\
&&+ \left(\ell\frac{d\rho}{\rho}-\ell[\lambda^+{+}\lambda^-]d\phi+[A^-{-}A^+]dt\right)^2 \  .
\end{eqnarray}
When the metric (\ref{compac}) is evaluated on Einstein's equations with a negative cosmological constant
\begin{equation}\label{eeom}
R_{\mu \nu}-\frac{1}{2}R g_{\mu\nu}-\frac{1}{\ell^2}g_{\mu\nu}=0 \ ,
\end{equation}
they reduce notably to temporal and angular derivatives of two independent copies of the AKNS nonlinear integrable equations,
\begin{subequations}\label{akns}
  \begin{align}\label{akns1}
   \pm \dot{r}^\pm+\frac{1}{\ell}\left(C'^\pm-2r^\pm A^\pm-2\lambda^\pm C^\pm\right)&=0 \ , \\ \label{akns2}
    \pm \dot{p}^\pm+\frac{1}{\ell} \left( B'^\pm+2p^\pm A^\pm+2\lambda^\pm B^\pm\right)&=0\ , \\ \label{akns3}
    A'^\pm-p^\pm C^\pm+r^\pm B^\pm&=0 \ .
  \end{align}
\end{subequations}
Here we denote the derivatives with respect to the angular and temporal variables as $\partial_\phi f \equiv  {f}^\prime$ and $\partial_t f \equiv\dot{f}$, respectively. While the first four equations determine the dynamics of the variables $r^\pm$ and $p^\pm$, the last two equations show that the functions $A^\pm$, $B^\pm$ and $C^\pm$ are not completely independent. As we will see below, all these equations allow to define the integrable hierarchies recursively.  To understand the connection between integrability and geometry in more detail, it is important to first review some crucial features of the integrable AKNS hierarchy, as they are now inherited by the solutions of General Relativity.
﻿
\subsection{A review on the AKNS integrable nonlinear hierarchy}\label{secakns}
﻿
Most of the well-known 1+1 nonlinear integrable equations, namely sine-Gordon, nonlinear Schr\"odinger (NLS), KdV, and mKdV, among many others, share the remarkable property that  they can be expressed in terms of a Lax pair or a zero-curvature condition \cite{bookakns}.  In both formulations, the integrable models are related to linear problems, which in several cases are related to quantum models. To see this, consider the set of two linear equations
\begin{equation}\label{linear}
\partial_{\phi}\Psi =U \Psi \ , \quad \partial_{t}\Psi =V \Psi \ ,
\end{equation}
where $U$ and $V$ are $2 \times 2$ matrices and $\Psi$ is a $2$-dimensional vector. Here, $t$ is the temporal coordinate, while $\phi$ is associated with spatial dependence, which can be a coordinate on the line or circle depending on the field's boundary conditions.  The compatibility condition $\partial_\phi\partial_t\Psi=\partial_t\partial_\phi\Psi$, together with the linear equations (\ref{linear}), lead to the zero curvature condition, 
\begin{equation}\label{comp}
\partial_t U-\partial_\phi V+ [U,V]=0 \ .
\end{equation}
All the aforementioned integrable systems can be unified in one single set of equations, the AKNS system, that encompasses them in terms of the following two-dimensional matrices \cite{bookakns},
 \begin{equation}\label{aphi}
U= \left(
    \begin{matrix}
   -  \lambda  & p(t,\phi) \\
      r(t,\phi) &  \lambda
    \end{matrix}
\right) \  ,\quad V= \left(
    \begin{matrix}
  A(t,\phi;\lambda)   & B(t,\phi;\lambda)  \\
     C(t,\phi;\lambda) & -A(t,\phi;\lambda)  
    \end{matrix}
\right) \ ,
\end{equation}
where $\lambda$ is the so-called spectral parameter, which generically is a complex variable. This choice of $U$ and $V$ allows to reduce the zero curvature condition (\ref{comp}) into three equations,
\begin{subequations}\label{aknsa}
  \begin{align}\label{aknsa1}
    \dot{r}-C'+2r A+2\lambda C&=0 \ , \\ \label{aknsa2}
     \dot{p}-B'-2p A-2\lambda B&=0\ , \\ \label{aknsa3}
    A'-p C+r B&=0 \ ,
  \end{align}
 \end{subequations}
 which have the same form as the chiral sector of the Einstein equations (\ref{akns}). The equations (\ref{aknsa}) become the AKNS hierarchy if we assume the functions $A,B$ and $C$ to be Laurent expansions in powers of the spectral parameter $\lambda$,
 \begin{equation}\label{rec_sol}
    A=\sum_{j=0}^{n} A_{j}\lambda^{n-j} \ , \quad  B=\sum_{j=0}^{n} B_{j}\lambda^{n-j}\ , \quad C=\sum_{j=0}^{n} C_{j}\lambda^{n-j} \ ,
\end{equation}
where $n$ is an arbitrary integer. This assumption defines a set of recursion relations,
\begin{subequations}\label{recur}
  \begin{align}
    A_{j}'&=qC_{j}-rB_{j} \ , \label{recurA}\\
    B_{j+1}&=-\frac{1}{2}B_{j}'-p A_{j}\ ,  \\
    C_{j+1}&=\frac{1}{2}C_{j}'-r A_{j}\ ,
  \end{align}
\end{subequations}
along with $B_0=C_0=0$ and  $A_{0}$ an arbitrary constant. The dynamical equations are reduced to
\begin{equation}\label{eomN}
 \dot r = C_n'- 2r A_n\ ,  \quad  \dot p = B_n'+2 p A_n \ .
\end{equation}
These set of equations \eqref{aknsa} are known as the $n$-th member of the hierarchy of AKNS equations \cite{gesztesy}. The recursion relation \eqref{recur} shows that the expansion's coefficients $A_j, B_j$, and $C_j$ are functionally related to the dynamical fields $p$ and $r$ and their derivatives. Consequently, the order of the angular derivatives appearing in the equation \eqref{eomN} increases for larger values of $n$.   
﻿
﻿To illustrate the mechanism of generating higher order equations, we start by choosing the case $n=3$ to solve the system \eqref{aknsa}. Using the expansion \eqref{rec_sol} with the initial conditions  $B_0=C_0=0, A_0=a_0$, this leads to the following equations,
\begin{subequations}\label{3order}
\begin{align}
\dot{p}&+ a_0(p'''-6prp')+a_1(p''-2p^2 r)+a_2p'-a_3 p=0 \  , \\ 
\dot{r}&+ a_0(r'''-6prr')-a_1(r''-2p r^2)+a_2r'+a_3 r=0  \  ,
\end{align}
\end{subequations}
where $a_i$, $i=0,1,2,3$ are integration constants which appear from the integration of the equation (\ref{recur}). The choice of specific values of these integration constants plus some conditions between $p$ and $r$ reduce the AKNS equations into familiar integrable examples:
\begin{itemize}
\item  {\bf Korteweg-de Vries (KdV)} Setting $r=-1$, $a_0=1$  and $a_1=a_2=a_3=0$,
\begin{equation}\label{ksta}
\dot{p}+p'''+6 p p'=0 \ .
\end{equation}
\item  {\bf Modified Korteweg-de Vries (mKdV)} Setting $r=p$, $a_0=1$ and $a_1=a_2=a_3=0$,
\begin{equation}\label{ksta}
\dot{p}+p'''+6 p^2 p'=0 \ .
\end{equation}
\item  {\bf Non-linear Schr\"odinger (NLS)}
Setting $r=-\kappa p^*$, $a_1=1$ and $a_0=a_2=a_3=0$,
\begin{equation}\label{ksta}
\dot{p}+p''- 2\kappa p |p|^2=0 \ ,
\end{equation}
where the positive (negative) value of $\kappa$ correspond to the defocusing (focusing) case.
\end{itemize}
﻿
﻿
As an integrable model with infinite degrees of freedom, the AKNS system has an infinite set of conserved charges $H_i$, which are in involution under suitable defined Poisson brackets,
\begin{equation}\label{aknsch}
\{H_i,H_j\}=0 \ , \quad i,j \in \mathbb{N}. 
\end{equation}
These conserved charges can be obtained recursively due to the bi-Hamiltonian structure of the AKNS equations. The first non-trivial charges are
\begin{equation}\label{AKNS-Cargas}
  H_{2}=\frac{1}{2}\int pr \ d\phi \ ,\quad H_{3}=\frac{1}{4}\int (p'r-pr')d\phi  \ ,\quad  H_{4}=-\frac{3}{8}\int (p^{2}r^{2}+p'r')d\phi \ .
\end{equation}
﻿
In section \ref{intsec}, we will offer more details about the bi-Hamiltonian structure, which is decisive in determining the commuting property of the gravitational conserved charges.
﻿
\subsection{Lax pairs and stationary hierarchies}
The two linear equations (\ref{linear}) can also be expressed in what is known as the Lax pair formulation, 
\begin{equation}\label{lax1}
L \psi = \lambda \psi \ , \quad  \psi_t= M \psi \ , 
\end{equation}
with the operators $L$ and $M$ and the auxiliary field $\psi$ and an arbitrary complex spectral parameter $\lambda$. The expressions for the operators can be obtained from (\ref{linear}), and the compatibility of the two linear equations (\ref{lax1}) is the famous Lax equation
\begin{equation}
\dot{L}+[L,M]=0 \ ,
\end{equation}
where the square brackets denote commutation. The main advantage of the Lax equation is that it immediately implies the conservation of infinitely many integrals of motion or conserved laws. Both the Lax pair and the zero curvature condition, i.e. the equations (\ref{linear}) and (\ref{lax1}), are equivalent frameworks where one switches from linear operators related to an eigenvalue equation for $\lambda$ to a matrix formulation, usually involving powers of the spectral parameter. The recursion relations in terms of polynomials of $\lambda$ in (\ref{rec_sol}) are exactly these powers in the parameter. If we consider now that the fields are time independent, $p(t,\phi)=p(\phi)$ and $r(t,\phi)=r(\phi)$ and the same for $A,B$ and $C$, we enter to the scheme known as the stationary AKNS hierarchy \cite{gesztesy}. In this scenario, the Lax equation is zero $[L,M]=0$ if $p$ and $r$ solve the $n$th stationary AKNS equations, which for $n=3$ are of the form (\ref{3order}) setting $\dot{p}=\dot{r}=0$. In this case, the commuting Lax operators $L$ and $M$ satisfy the Burchnall-Chaundy theorem, implying that there exists an algebraic relation, in this case a polynomial, between these two operators, explicilty
\begin{equation}\label{spec}
[L,M]=0  \quad \longrightarrow \quad M^2=R_{2n+2}(L)=\prod_{m=0}^{2n+2}(L-E_m) \ ,
\end{equation}
where $E_m$ are some constants that depend on the solutions $p$ and $r$. The polynomial $\prod_{m=0}^{2n+2}(L-E_m)$ is nothing else than the so-called spectral curve or spectral polynomial. Making the substitution $y=L$ and $M=\lambda$ transform the equation (\ref{spec}) in a hyperelliptic curve of genus $n$ which can be expressed exactly in terms of the recursive functions $A,B$ and $C$,
\begin{equation}\label{spec2}
y^2=\prod_{m=0}^{2n+2}(\lambda-E_m)=A_n^2+B_nC_n \ ,
\end{equation}
This remarkable relationship between the hyperelliptic curve and the Lax operators implies that the right-hand side of \eqref{spec2} remains constant, thereby establishing a deep connection between integrability and algebraic geometry \cite{gesztesy}. 
In special cases such as KdV, mKdV, or NLS, the spectral polynomials acquire a physical interpretation in terms of the eigenvalues of finite-gap systems arising in associated quantum problems. In this context, the constants $E_m$ correspond to singlet band-edge states that reside at the boundaries of the allowed energy bands \cite{Correa:2008hc, Correa:2008bz}. As we will see below, all of these features are relevant to the black hole properties that underlie the geometric integrable structure.
﻿
\subsection{Chern-Simons gravity, the zero curvature equation and the AKNS integrable system}
The embedding of AKNS equations within three-dimensional gravity with negative cosmological constant is most simply achieved in the Chern-Simons formulation \cite{cs1, cs2}. In such an approach,  Einstein's equations of motion (\ref{eeom}) can be recast as zero curvature conditions associated with two $sl(2,\mathbb{R})$ gauge fields $\mathcal{A}^\pm$, 
\begin{equation}\label{zcs}
\mathcal{F}^{\pm}=0 \  , \quad \mathcal{F}^{\pm}=d\mathcal{A}^\pm+\mathcal{A}^\pm \wedge \mathcal{A}^\pm \  .
\end{equation}
﻿
The $sl\left(2,\mathbb{R}\right)$ algebra is spanned by the generators $L_{n}, \ n\in\{-1,0,1\}$, satisfying $\left[L_{n},L_{m}\right]=\left(n-m\right)L_{n+m}$. We are using the matrix representation,
\begin{equation}\label{matrix-rep}
L_{-1}=\begin{pmatrix}
    0  & 0 \\
    1&0     
\end{pmatrix},
\qquad
L_{0}=\begin{pmatrix}
     -\tfrac{1}{2}  & 0 \\
      0 &\tfrac{1}{2}    
\end{pmatrix},\qquad
L_{1}=\begin{pmatrix}
    0 & -1 \\
      0& 0    
\end{pmatrix},
\end{equation}
associated to the nonvanishing components of the invariant bilinear form are $\langle L_{1},L_{-1}\rangle=-1$ and $\langle L_{0}, L_{0}\rangle=1/2$. The relationship between $\mathcal{A}^\pm$  and the metric field is given by
\begin{equation}\label{gmunu}
  g_{\mu\nu}=\frac{\ell^2}{2}\left\langle \left(\mathcal{A}_{\mu}^{+}-\mathcal{A}_{\mu}^{-}\right),\left(\mathcal{A}_{\nu}^{+}-\mathcal{A}_{\nu}^{-}\right)\right\rangle \  .
\end{equation} 
﻿
﻿
In the Chern-Simons approach,  the metric boundary conditions are constructed by choosing an appropriate boundary behavior for the gauge fields $\mathcal{A}^\pm$. To recover (\ref{metricaAKNS}), a first gauge choice is used to capture the radial dependence \cite{Coussaert:1995zp},
\begin{equation}\label{bdab}
\mathcal{A}^\pm=b^{\mp 1}( d+ a^\pm ) b^{\pm 1} \  ,
\end{equation}
where the gauge group element $b$ is chosen as $b\left(\rho\right)=\exp[\log(\rho/\ell)L_{0}]$,  thus rendering an asymptotically locally AdS spacetime. The auxiliary connections $a^\pm=a^{\pm}_{\phi}d\phi+a^{\pm}_{t}dt$ are functions only of $t$ and $\phi$, while the zero curvature conditions (\ref{zcs}) reduce to
 \begin{equation}\label{zcs2}
\partial_t a^{\pm}_{\phi}-\partial_\phi a^{\pm}_{t}- [a^{\pm}_{\phi},a^{\pm}_{t}]=0 \  .
\end{equation}
The above equations become the dynamics of two copies of the AKNS system \eqref{akns} with opposite chirality after considering
\begin{align}
& a_\phi^{+}=\begin{pmatrix}
    \lambda^{+}\  & p^{+} \\
      r^{+} &- \lambda^{+}     
\end{pmatrix},
\qquad\ \ \ \ \ 
a_\phi^{-}=\begin{pmatrix}
     -\lambda^{-}\  & -r^{-} \\
      -p^{-} & \lambda^{-}     
            \end{pmatrix},
\\
& a_t^{+}=\frac{1}{\ell}\begin{pmatrix}
    A^{+}\  & -B^{+} \\
      -C^{+} &- A^{+}     
\end{pmatrix},
\qquad
a_t^{-}=\frac{1}{\ell}\begin{pmatrix}
     A^{-}\  & -C^{-} \\
      -B^{-} & -A^{-}    
\end{pmatrix}.
\end{align}
In terms of the algebra generators, the auxiliary gauge fields are given by
\begin{equation}
a_{\phi}^{\pm}=\mp 2\lambda^{\pm}L_{0}-p^{\pm}L_{\pm 1}+r^{\pm}L_{\mp},\qquad a_{t}^{\pm}=\frac{1}{\ell}(-2A^{\pm}L_{0}\pm B^{\pm}L_{\pm 1} \mp C^{\pm}L_{\mp 1}).
\end{equation}
﻿
Finally,  the comparison between the AKNS matrices \eqref{aphi},  and the above gauge field matrices is given by the following identifications,
\begin{eqnarray}
 & a^{+}_\phi\rightarrow -4L_{0}U L_{0} \ ,\qquad  &  a_\phi^{-}\rightarrow 4L_{0}U^{T}L_{0} \ , \\
 & a^{+}_{t}\rightarrow \displaystyle\frac{4}{\ell}L_{0}V L_{0} \ , \qquad  & a_{t}^{-}\rightarrow    \frac{4}{\ell}L_{0}V^{T}L_{0} \ ,
\end{eqnarray}
where here $^T$ stands for transposition.
﻿
﻿
﻿
﻿
\section{Asymptotic structure}
﻿
﻿
This section shows that the infinite set of commuting integrals of motion associated with the AKNS integrable system generates the asymptotic symmetries of the spacetime \eqref{metricaAKNS}.  Using the  Regge-Teitelboim Hamiltonian approach for the generic metric fall-off \eqref{asym1}, we show that the AKNS integrals of motion correspond to the gravitational canonical charges, provided suitable integrability conditions for the gauge parameters. We introduce the bi-Hamiltonian structure of AKNS to prove the abelian nature of the charge algebra and show that energy and angular momentum are contained within the set of asymptotic charges.
﻿
﻿
﻿
﻿
\subsection{AKNS symmetries}
﻿
﻿
 We search for diffeomorphisms that preserve the form of the metric  \eqref{metricaAKNS}. This means to find vectors $\eta$ that map gravitational configurations into themselves $\delta_{\eta} g_{\mu\nu}   = \mathcal{O}( g_{\mu\nu})$.  In this case, $\eta$ was constructed using the dictionary among the metric and the Chern-Simons formulation, which relates gauge transformations and symmetries generated by the vector $\eta$,  up to terms proportional to the field equations. See appendix \ref{Appendix A} for details. Indeed, we solve the Killing equation, 
\begin{equation} \label{variation}
\delta_{\eta} g_{\mu\nu}   = \mathcal{L}_\eta g_{\mu\nu}+ \text{eom}, 
\end{equation}
where the exact form of the vector components $\eta=\eta^{t}\partial_t + \eta^{\rho} \partial_\rho+\eta^{\phi}\partial_\phi$ is,
\begin{subequations}﻿ \label{killingss}
\begin{align}\label{killing1}
\eta^{t} & =\ell\frac{(\rho^2 p^-+\ell^2 r^+)(\ell^2 \gamma^{-}-\rho^2 \beta^+)+(\rho^2 p^++\ell^2 r^-)(\ell^2 \gamma^{+}-\rho^2 \beta^-)}{(\rho^2 p^-+\ell^2 r^+)(\rho^2 B^+-\ell^2 C^-)+(\rho^2 p^++\ell^2 r^-)(\rho^2 B^--\ell^2 C^+)} \  ,\\ \label{killing2}
\eta^{\rho} & =\alpha^+-\alpha^-+\frac{(A^{-}-A^+)}{\ell}\eta^{t}-(\lambda^++\lambda^-)\eta^{\phi} \  , \\ \label{killing3}
\eta^{\phi} & =\frac{(\rho^2 B^--\ell^2 C^+)(\ell^2 \gamma^--\rho^2 \beta^+)+(\rho^2 B^+-\ell^2 C^-)(\rho^2 \beta^--\ell^2 \gamma^+)}{(\rho^2 B^+-\ell^2 C^-)(\rho^2 p^-+\ell^2 r^+)+(\rho^2 B^--\ell^2 C^+)(\rho^2 p^++\ell^2 r^-)} \  .
\end{align}
\end{subequations}
The functions $\alpha^\pm,\beta^\pm$, and $\gamma^\pm$ depend generically on $t$ and $\phi$, and they characterize the asymptotic symmetries with non-trivial canonical charges. Invariance of \eqref{metricaAKNS} under transformation \eqref{variation} also implies the additional relation
\begin{equation} \label{consistency_param}
\alpha^{\pm\prime}=\gamma^{\pm}p^{\pm}-\beta^{\pm}r^{\pm},
\end{equation}
resembling equation \eqref{aknsa3}. In subsection \ref{Symmetry transformations}
we use the above equation to show that $\alpha^\pm,\beta^\pm$ and $\gamma^\pm$ are state-dependent functions,  which will then further be used to construct integrability conditions for the canonical generators associated with the AKNS charges. 
﻿
The condition \eqref{variation} also introduces the asymptotic symmetry transformations on the  dynamical fields,
\begin{subequations}\label{field-transform}
\begin{align}
  \delta p^{\pm}&=\mp2p^{\pm}\alpha^{\pm}\mp\lambda^{\pm}\beta^{\pm}\mp\beta^{\pm \prime}\ ,\\
  \delta r^{\pm}&=\pm2r^{\pm}\alpha^{\pm}\pm\lambda^{\pm}\gamma^{\pm}\mp\gamma^{\pm\prime}\ ,
\end{align}
\end{subequations}
and the transformations of the functions $A^\pm$, $B^\pm$ and $C^\pm$, 
\begin{subequations}\label{lagrange-transform}
\begin{eqnarray}
\delta A^{\pm}&=\pm C^{\pm}\beta^{\pm}\mp B^{\pm}\gamma^{\pm}+\ell\dot{\alpha}^{\pm},\\
\delta B^{\pm}&=\mp2B^{\pm}\alpha^{\pm}\pm A^{\pm}\beta^{\pm}+\ell\dot{\beta}^{\pm},\\
\delta C^{\pm}&=\pm2C^{\pm}\alpha^{\pm}\mp A^{\pm}\beta^{\pm}+\ell\dot{\gamma}^{\pm}.
\end{eqnarray}
\end{subequations}
﻿
﻿
﻿
The asymptotic expansions of the vector components (\ref{killing1})-(\ref{killing3}) in the radial coordinate have the form,
\begin{align}\label{asym killing1}
\eta^{t} & =Y+\frac{\overline{Y}}{\rho^{2}}+\frac{\overline{\overline{Y}}}{\rho^{4}}+\mathcal{O}\left(\frac{1}{\rho^{6}}\right)\  ,\\ \label{asym killing2}
\eta^{\rho} & =\rho \Theta+\frac{\overline{\Theta}}{\rho}+\frac{\overline{\overline{\Theta}}}{\rho^3}+\mathcal{O}\left(\frac{1}{\rho^{5}}\right)\  ,\\\label{asym killing3}
\eta^{\phi} & =\Sigma +\frac{\overline{\Sigma}}{\rho^{2}}+ \frac{\overline{\overline{\Sigma}}}{\rho^{4}}+\mathcal{O}\left(\frac{1}{\rho^{6}}\right)\  .
\end{align}
where its radial fall-off deviates from the one observed in the Brown-Henneaux case, where $\overline{\overline{Y}},\overline{\overline{\Theta}}$ and $\overline{\overline{\Sigma}}$ are new contributions. This is naturally expected from the metric behavior (\ref{metricaAKNS}) since it requires a relaxed contribution in the expansion in each of the components of the Killing vectors in order to ensure the full realization of the AKNS asymptotic symmetry. The asymptotic functions can be written as
\begin{eqnarray*}
Y/\ell & = &-  \frac{\vec{p}\cdot \cev{\beta}}{\vec{p}\cdot \cev{B}}= -\frac{p^{-}\beta^{+}{+}p^{+}\beta^{-}}{p^{-}B^{+}{+}p^{+}B^{-}}  \  ,\\
\overline{Y}/\ell^3 & = & \frac{(\vec{p}\cdot \cev{\beta})(\vec{r}\cdot \vec{B}-\vec{p}\cdot \vec{C})}{(\vec{p}\cdot \cev{B})^{2}} +\frac{\vec{p}\cdot \vec{\gamma}-\vec{r}\cdot \vec{\beta}}{\vec{p}\cdot \cev{B}}\  ,\\
\overline{\overline{Y}}/\ell^5 & = &-\frac{(\vec{p}\cdot \cev{\beta})(\vec{p}\cdot \vec{C}{-}\vec{r}\cdot \vec{B})^2}{(\vec{p}\cdot \cev{B})^{3}}{+}\frac{(\vec{r}\cdot \vec{B}{-}\vec{p}\cdot \vec{C})(\vec{r}\cdot \vec{\beta}{-}\vec{p}\cdot \vec{\gamma}){+}(\vec{r}\cdot \cev{C})(\vec{p}\cdot \cev{B})}{(\vec{p}\cdot \cev{B})^{2}}{+}\frac{\vec{r}\cdot \cev{\gamma}}{\vec{p}\cdot \cev{B}}\  ,\\
\overline{\Sigma}/\ell^2 & = & \frac{(\vec{B}\ast \cev{\beta})(\vec{r}\cdot \vec{B}-\vec{p}\cdot \vec{C})}{(\vec{p}\cdot \cev{B})^{2}} +\frac{\vec{C}\ast \vec{\gamma}-\vec{C}\ast  \vec{\beta}}{\vec{p}\cdot \cev{B}}\  ,\\
\overline{\overline{\Sigma}}/\ell^4 & = &-\frac{(\vec{B}\ast \cev{\beta})(\vec{p}\cdot \vec{C}{-}\vec{r}\cdot \vec{B})^2}{(\vec{p}\cdot \cev{B})^{3}} 
{+}\frac{(\vec{r}\cdot \vec{B}{-}\vec{p}\cdot \vec{C})(\vec{C}\ast \vec{\beta}{-}\vec{B}\ast \vec{\gamma}){+}(\vec{B}\ast \cev{\beta})(\vec{r}\cdot \vec{C})}{(\vec{p}\cdot \cev{B})^{2}}{+}\frac{\vec{C}\ast \cev{\gamma}}{\vec{p}\cdot \cev{B}}\  ,
\end{eqnarray*}
﻿
\begin{eqnarray*}
\Theta & = & \left(\alpha^{+}-\alpha^{-}\right)+\frac{\left(A^{-}-A^{+}\right)}{\ell}Y-(\lambda^{+}+\lambda^{-})\  ,\\
\overline{\Theta} & = & \frac{\left(A^{-}-A^{+}\right)}{\ell}\overline{Y}-(\lambda^{+}+\lambda^{-})\overline{\Sigma} \  ,
\\
\overline{\overline{\Theta}} & = & \frac{\left(A^{-}-A^{+}\right)}{\ell}\overline{\overline{Y}}-(\lambda^{+}+\lambda^{-})\overline{\overline{\Sigma}} \  .
\end{eqnarray*}
Here we have introduced the notation $\vec{x}=(x^{+},x^{-})$ and $\cev{x}=(x^{-},x^{+})$ with the two products; $\vec{x}\cdot \vec{y}=x^+y^+{+}x^-y^-$, $ \cev{x}\cdot \vec{y}=\vec{x}\cdot \cev{y}=x^+y^-{+}x^-y^-$ as the standard dot product and $\vec{x}\ast \vec{y}=x^+y^+{-}x^-y^-$ and $\cev{x}\ast \vec{y}=x^+y^--x^-y^-$.  
﻿
It is worth noting that applying diffeomorphisms generated by the vector $\eta$ modifies the $1/\rho$ coefficient in the $g_{\rho\phi}$ metric component, rendering it non-constant. However, this modification vanishes when the equations of motion \eqref{akns} are satisfied, along with the corresponding transformations of the fields \eqref{field-transform} and \eqref{lagrange-transform}. The above derivation is consistent with the results obtained in the gauge formulation of the theory \cite{cclp}. For more details see the Appendix \ref{Appendix A}.﻿
﻿
﻿
\subsection{Canonical charges}
﻿
﻿
﻿
﻿
﻿
Having found in the previous section the vector that preserves the form of the AKNS metric, we turn to calculate its corresponding conserved charges. We consider the generic asymptotic behavior \eqref{asym1} and follow the Regge-Teitelboim Hamiltonian approach \cite{Regge:1974zd}. For the latter, let us consider the canonical pair $(g_{ij},\pi^{ij})$, where $g_{ij}$ stands for the metric of the space-like hypersurface associated with the 2+1 decomposition of the metric, 
\begin{equation}\label{admetric}
ds^2=-N^2 dt^2 + (N^i dt +dx^i)(N^j dt + dx^j)g_{ij},
\end{equation}
where $N^{\perp}$ and $ N^{i}$ are the lapse and shift functions, respectively. The Hamiltonian approach provides the functional variation of the canonical generator $Q[\epsilon^{\perp},\epsilon^{i}]$,  given by the surface integral,
\begin{equation} \label{deltaQ hamiltonian}
    \delta Q[\epsilon^{\perp},\epsilon^{i}]=\frac{k}{2\pi}\int dS_l \left[ G^{ijkl}\left(\epsilon^{\perp}\nabla_k\delta g_{ij}-\nabla_{k}\epsilon^{\perp}\delta g_{ij}\right)+2\epsilon_{k}\delta \pi^{kl}+ (2\epsilon^j\pi^{kl}-\epsilon^l\pi^{jk})\delta g_{jk} \right],
\end{equation}
where $G^{ijkl}=\tfrac{\sqrt{g}}{2}(g^{ik}g^{jl}+g^{il}g^{jk}-2g^{ij}g^{kl})$, $k$ is related with the AdS radius and the gravitational constant $G$ by the equation $k=\tfrac{\ell}{4G}$. The corresponding asymptotic Killing vectors  $\eta^{\mu}=\eta^t \partial_t+\eta^i\partial_x^i$  are related to the surface deformation parameters $\epsilon^{\perp}$ and $\epsilon^{i}$ through,
\begin{equation}\label{epsilonN}
\epsilon^{\perp}=N\ \eta^t\qquad\epsilon^{i}=\eta^i+N^{i}\eta^t.
\end{equation}
﻿
﻿
﻿
﻿
﻿
﻿
﻿
The asymptotic behavior of the canonical pair $(g_{ij},\pi^{ij})$ contained within the asymptotic conditions \eqref{asym1} is given by
\begin{subequations}\label{spatial metric}
  \begin{align} 
    g_{\rho\rho}	&=\frac{\ell^{2}}{\rho^{2}}+\mathcal{O}\left(\frac{1}{\rho^{4}}\right),\\
    g_{\rho\phi}	&=\frac{g_{\rho\phi}^{(-1)}}{\rho}+\mathcal{O}\left(\frac{1}{\rho^{2}}\right),\\
    g_{\phi\phi}	&=g_{\phi\phi}^{(2)}\rho^{2}+g_{\phi\phi}^{(0)}+g_{\phi\phi}^{(-2)}\rho^{-2}+\mathcal{O}\left(\frac{1}{\rho^3}\right),
  \end{align}
\end{subequations}
while the momenta behaves as
\begin{subequations}\label{momenta}
  \begin{align}
    \pi^{rr}&=h^{rr}\left(t,\phi\right)\rho+\mathcal{O}\left(\frac{1}{\rho}\right),\\
    \pi^{r\phi}&=\frac{h^{r\phi}\left(t,\phi\right)}{\rho^{2}}+\mathcal{O}\left(\frac{1}{\rho^{4}}\right),\\
    \pi^{\phi\phi}&=\mathcal{O}\left(\frac{1}{\rho^{5}}\right).
  \end{align}
\end{subequations}
﻿Under these asymptotic conditions, the constraints behave as 
\begin{equation}\label{constraints}
H^{\perp}=\mathcal{O}\left(\frac{1}{\rho^{3}}\right) \qquad H^{\rho}=\mathcal{O}\left(\frac{1}{\rho^{3}}\right)\qquad H^{\phi}=\mathcal{O}\left(\frac{1}{\rho}\right) \ .
\end{equation}
Furthermore, the asymptotic expansion for the lapse and shift functions corresponds to
\begin{equation}
  N=\mathcal{O}\left(\rho\right),\quad N^{\rho}=\mathcal{O}\left(\rho\right), \quad N^{\varphi}=\mathcal{O}\left(1\right).
\end{equation}
Hence, when considering the vector \eqref{asym killing1} and the relation \eqref{epsilonN}, the surface deformation parameters behaves asymptotically as
\begin{equation}\label{surface asympt}
\epsilon^{\perp}=\mathcal{\epsilon}_{(0)}^{\perp}\rho+\frac{\mathcal{\epsilon}_{(-1)}^{\perp}}{\rho}+\mathcal{O}\left(\rho^{-2}\right),\quad\epsilon^{\rho}=\mathcal{\epsilon}_{(1)}^{\text{\ensuremath{\rho}}}\rho+\mathcal{O}\left(\rho^{0}\right),\quad\epsilon^{\phi}=\mathcal{\epsilon}_{(0)}^{\phi}+\mathcal{O}\left(\rho^{-1}\right),
\end{equation}
where the coefficients $\mathcal{\epsilon}_{(0)}^{\perp},\mathcal{\epsilon}_{(-1)}^{\perp},\mathcal{\epsilon}_{(1)}^{\text{\ensuremath{\rho}}},\mathcal{\epsilon}_{(0)}^{\phi}$ are functions of $t$ and $\phi$.
﻿
﻿
 
﻿
We verify that the proposed asymptotic behavior for the gravitational fields \eqref{spatial metric},\eqref{momenta} and the surface deformation generators \eqref{surface asympt} leads to 
\begin{eqnarray}\label{charge variation}
\nonumber \delta Q&=&\frac{k}{2\pi}\int\displaylimits_{\rho\rightarrow \infty} d\phi\Bigg(\mathcal{\epsilon}_{(0)}^{\perp}\delta\left[\frac{g_{\phi\phi}^{(0)}-g_{\rho\phi}^{(-1)}}{\ell\sqrt{g_{\phi\phi}^{(2)}}}\right]-\epsilon_{(-1)}^{\perp}\frac{2}{\ell}\delta\left(\sqrt{g_{\phi\phi}^{(2)}}\right) \\
 &&\qquad\qquad\qquad+ 2\mathcal{\epsilon}_{(0)}^{\phi}\delta\left(g_{\rho\phi}^{(-1)}h^{rr}+g_{\phi\phi}^{(2)}h^{r\phi}\right)+2\ell^{2}\mathcal{\epsilon}_{(1)}^{\text{\ensuremath{\rho}}}\delta h^{rr}\Bigg).
\end{eqnarray}
The above expression shows that proposed asymptotic behavior is enough to make the surface integral  \eqref{deltaQ hamiltonian}  finite when evaluated at infinity\footnote{Here we have considered $\delta g_{\rho\phi}^{(-1)}=0$, to make contact with the AKNS asymptotic conditions \eqref{metricaAKNS}.}.
To specify to the AKNS metric \eqref{metricaAKNS}, we identify the lapse and shift components of (\ref{admetric}) as
\begin{align}\label{lapse}
  N^2&=\frac{\rho^2}{4\ell^2}\frac{\left(    \Omega^+   \omega^-  + \Omega^- \omega^+   \right)^2}{   \omega^- \omega^+  } \ , \\ \label{shift}
    N^\rho&=\frac{\rho}{\ell} \left[A^- - A^+  + \frac{1}{2} \left(\lambda^+ + \lambda^-        \right)    \left(\frac{\Omega^-}{\omega^-}-\frac{\Omega^+}{\omega^+}\right)\right] \ ,\\
 N^\phi&=\frac{1}{2\ell}  \left(\frac{\Omega^-}{\omega^-}-\frac{\Omega^+}{\omega^+}\right) ,
 \end{align}
 where the auxiliary functions $\Omega^\pm$ and $\omega^\pm$ are given by, 
\begin{equation}\label{omegas}
  \Omega^\pm\equiv B^\pm - \frac{\ell^2}{\rho^2}C^\mp \ , \quad   \omega^\pm\equiv p^\pm + \frac{\ell^2}{\rho^2}r^\mp \ .
\end{equation}
 Furthermore, the spatial part of the metric (\ref{admetric}) is
 \begin{equation}\label{gij}
  g_{ij}= \left(
    \begin{matrix}
      \frac{\ell^2}{\rho^2}& - \frac{\ell^2}{\rho}\left(\lambda^++\lambda^- \right)\\
       - \frac{\ell^2}{\rho}\left(\lambda^++\lambda^- \right)& \ell^2 \left(\lambda^++\lambda^- \right)^2+\rho^2 \omega^-\omega^+
    \end{matrix}
\right) \ .
\end{equation}
Using the above information, we find that the variation of the canonical charge \eqref{deltaQ hamiltonian} is given by  
 \begin{equation}\label{deltaQ}
\delta Q[\eta]= -\frac{k}{2\pi}\int\displaylimits d \phi \left( \beta^{+} \delta r^{+}+\gamma^{+} \delta p^{+}+\beta^{-} \delta r^{-}+\gamma^{-} \delta p^{-}\right).
\end{equation}
﻿
We find that the charge is finite but a priori non-integrable. It includes all four different varying dynamical fields $p^{\pm}, r^{\pm}$ and their corresponding conjugate functions $\beta^{\pm}$ and $\gamma^{\pm}$, consistent with the result obtained using the Chern-Simons formulation. However, unlike such an approach, the equations \eqref{akns3} and its variations, as well as \eqref{consistency_param}, must be enforced.

﻿
﻿
\subsubsection{Symmetry transformations,  {\it integrability} conditions}\label{Symmetry transformations}
﻿
﻿
To solve the {\it integrability} problem,  we follow the same scheme that generates the AKNS hierarchy. Similarly to (\ref{rec_sol}),  we start by assuming the expansion,
\begin{equation}\label{rec_solaa}
    \alpha^{\pm}=\sum_{n=0}^{M} \alpha^{\pm}_{n}(\lambda^{\pm})^{ M-n} \ , \quad  \beta^{\pm}=\sum_{n=0}^{M} \beta^{\pm}_{n}(\lambda^{\pm})^{ M-n} \ , \quad \gamma^{\pm}=\sum_{n=0}^{M} \gamma^{\pm}_{n}(\lambda^{\pm})^{ M-n} \ ,
\end{equation}
where $M$ is an arbitrary positive integer. In what follows, this assumption will show that the symmetry transformations \eqref{field-transform} and \eqref{lagrange-transform}, when they are properly treated,  unfold two types of information.  On the one hand,  they provide recursion relations to generate the coefficients of \eqref{rec_solaa}, in the same way as in \eqref{rec_sol}, which are tide up to the {\it integrability} conditions necessary to functionally integrate \eqref{deltaQ}. On the other hand,  they generate the symmetry transformations of the fields that will realize the AKNS charges.  
﻿In this last regard,  we evaluate \eqref{rec_solaa} on the gauge transformations \eqref{field-transform},  so that to find that the coefficients of the above expansion can be decoupled, as they appear associated to certain orders of $\lambda^\pm$. In particular, at order $(\lambda^{\pm})^{0}$, the resulting equations are 
\begin{equation}\label{eomM}
  \pm\delta r^{\pm} = -\gamma_M'+2 r^{\pm} \alpha_M\ , \quad {\pm}\delta p^{\pm} = -\beta^{\pm \prime}_M-2 p^{\pm} \alpha^{\pm}_M \ ,
\end{equation}
which are the actual field transformation laws of the dynamical fields,  only generated by the $M$-th coefficient of the expansion. Then, the integer $M$ also determines the dimension of the asymptotic symmetries. The remaining terms can be summarized as follows,
\begin{equation}\label{recurrence}
\begin{aligned}
 \beta^{\pm}_{n+1}&=-\frac{1}{2}\beta^{\pm \prime}_{n}-p^{\pm} \alpha^{\pm}_{n} \ , \\
\gamma^{\pm}_{n+1}&=\frac{1}{2}\gamma^{\pm \prime}_{n}-r^{\pm} \alpha^{\pm}_{n} \ ,
\end{aligned} \qquad\text{for}\quad 0\leq n \leq M-1
\end{equation}
along with $\beta^{\pm}_0=\gamma^{\pm}_0=0$. Applying the same expansion to the constraint equation \eqref{consistency_param}, we find that $\alpha^{\pm \prime}_{0}=0$ and
\begin{equation}\label{recurrenceA}
\alpha^{\pm \prime}_{n}=p^{\pm}\gamma^{\pm}_{n}-r^{\pm}\beta^{\pm}_{n},\qquad\text{for}\quad \quad 1\leq n \leq M \ .
\end{equation}
﻿All above recursion relations are crucial, as they allow to construct the variational relations \cite{Tu:1989},
\begin{equation}\label{functional rel}
 \alpha_{n}^\pm \equiv \frac{(n-1)}{2} \mathcal{H}_n^\pm \ , \quad \beta_{n-1}^\pm \equiv \frac{\delta \mathcal{H}_n^\pm}{\delta r^\pm} , \quad \gamma_{n-1}^\pm \equiv \frac{\delta \mathcal{H}_n^\pm}{\delta p^\pm}.  
\end{equation}
where $\mathcal{H}_{n}$ is Hamiltonian density $H_{n}=\int d\phi\mathcal{H}_{n}$ and $H_{n}$ is the $n-$th  AKNS conserved charge. Considering  \eqref{deltaQ}, we have,  
 \begin{eqnarray}
  \delta Q_{n}[\eta]&=&-\frac{k}{2 \pi}\left(\delta H^{+}_{n}+ \delta H^{-}_{n}\right)\\
 &=&-\frac{k}{2 \pi}\int d \phi \left( \beta_{n}^{+} \delta r^{+}+\gamma_{n}^{+} \delta p^{-}+\beta^{-} \delta r^{-}+\gamma_{n}^{-} \delta p^{-}\right) \ ,
\end{eqnarray}
and evaluate the functional relations \eqref{functional rel} to obtain $Q_n$ in terms of the AKNS charges \eqref{aknsch} and \eqref{AKNS-Cargas}.
In a more general form, evaluating the expansion \eqref{rec_solaa} and the integrability conditions \eqref{functional rel} in \eqref{deltaQ} leads to
\begin{equation}\label{Q}
 Q[\eta]=-\frac{k}{2 \pi}\left( \sum_{n=0}^M (\lambda^{+})^{M-n}H^{+}_{n+1}+ \sum_{n=0}^M (\lambda^{-})^{M-n}H^{-}_{n+1}\right) \ , 
\end{equation}
that realize the total charge as the sum of two terms weighted by the corresponding powers of the spectral parameters  $\lambda^\pm$.  The energy and angular momentum are included in the boundary conditions and can be obtained form the global charges, as generated by $\partial_{t}$ and $\partial_{\phi}$ and imposing extra conditions to the gauge parameters. In the case of the energy of the system $E=Q[\partial_{t}]$,
\begin{equation}\label{Energy}
\begin{aligned}
\alpha^\pm &=-A^\pm\\
\beta^\pm &=-B^\pm \\
\gamma^\pm &=-C^\pm  
\end{aligned} \quad \longrightarrow \quad \delta E=\frac{k}{2\pi}\int d \phi \left( B^{+} \delta r^{+}+C^{+} \delta p^{+}+B^{-} \delta r^{-}+C^{-} \delta p^{-}\right) \ ,
\end{equation}
which provided \eqref{rec_sol},  leads to
\begin{equation}
E=\frac{k}{2\pi}\sum_{n=0}^{N}(\lambda^{+})^{N-n}H^{+}_{n+1}+\frac{k}{2\pi}\sum_{n=0}^{N}(\lambda^{-})^{N-n}H^{-}_{n+1} \ .
\end{equation}
Note that the index $N$ in the sum is fixed by the dynamics of the system and will therefore contribute a \textit{finite} number of terms to the $E$, so that to make a distinction with the sum index $M$ in \eqref{Q}, labelling an \textit{infinite} but numerable set of terms. A special example of this will be shown in the section on black holes \ref{KDV black hole} using the stationary KdV case. In the case of the angular momentum $J=Q[\partial_{\phi}]$, it is obtained with an even more restricted choice of gauge parameters
\begin{equation}\label{Momentum}
\begin{aligned}
\alpha^\pm &=\pm \lambda^\pm\\
\beta^\pm &=\mp p^\pm \\
\gamma^\pm &=\mp r^\pm  
\end{aligned} \quad \longrightarrow \quad  \delta J=\frac{k}{2\pi} \int d \phi \left(p^{+} \delta r^{+}+r^{+} \delta p^{+}-p^{-} \delta r^{-}-r^{-} \delta p^{-}\right) \ ,
\end{equation}
whose integration reads 
 \begin{equation}\label{Momentum2}
 J=\frac{k}{2\pi} \int d \phi \left(p^{+}  r^{+}- r^{-}  p^{-}\right)= \frac{k}{\pi}(H^{+}_{2}-H^{-}_{2}) \ .
 \end{equation}
﻿
﻿
\subsection{Dynamical systems, Poisson structure and asymptotic symmetry algebras}\label{intsec}
﻿
The bi-Hamiltonian structure of the AKNS system allows the construction of two operators that independently generate the dynamics of the system. On the one hand, the AKNS dynamical equations can be obtained in two different ways by means of
\begin{equation}
\left(\begin{array}{c}
\dot{r}\\
\dot{p}
\end{array}\right)=\mathcal{D}_{1}\left(\begin{array}{c}
\mathcal{R}_{N+1}\\
\mathcal{P}_{N+1}
\end{array}\right)=\mathcal{D}_{2}\left(\begin{array}{c}
\mathcal{R}_{N+2}\\
\mathcal{P}_{N+2}
\end{array}\right) \ , 
\end{equation}
where the Hamiltonian operators are
\begin{equation}
\mathcal{D}_{1}=\frac{1}{\ell}\left(\begin{array}{cc}
-2r\partial_{\phi}^{-1}\left(r\cdot\right) & \quad-\partial_{\phi}+2r\partial_{\phi}^{-1}\left(p\cdot\right)\\
-\partial_{\phi}+2r\partial_{\phi}^{-1}\left(p\cdot\right) & 2p\partial_{\phi}^{-1}\left(p\cdot\right)
\end{array}\right) \ , \quad 
\mathcal{D}_{2}=\frac{1}{\ell}\left(\begin{array}{cc}
0 & \quad-2\\
2 & 0
\end{array}\right) \ ,
\end{equation}
and the antiderivative is defined as $\partial_{\phi}^{-1}f(\phi)=\int_{\infty}^{\phi}\left(f(y)\right)dy$ where the value of the integrand is set to zero at infinity.
One can choose any of the operators $\mathcal{\mathcal{D}}_{1}$ or $\mathcal{\mathcal{D}}_{2}$ to define a Poisson bracket structure, but they also allow to define the AKNS equations recursively,
\begin{equation}\label{bihamrec}
  \left(
    \begin{aligned}
      \mathcal{R}_{N+2}\\ \mathcal{P}_{N+2}
    \end{aligned}
\right) = \mathcal{D}_2^{-1}\mathcal{D}_1 \left(
    \begin{aligned}
      \mathcal{R}_{N+1}\\ \mathcal{P}_{N+1}
    \end{aligned}
\right) \ .
\end{equation}
The case for the operator $\mathcal{\mathcal{D}}_{1}$ leads to write the Poisson
brackets of two arbitrary functionals $F=F\left[r,p\right]$ and $G=G[r,p]$ in the following form, 
\begin{equation}\renewcommand{\arraystretch}{1.3}
\left\{ F\left(\phi\right),G\left(\phi'\right)\right\} =\int d\phi\left(\begin{array}{cc}
\frac{\delta F\left(\phi\right)}{\delta r\left(\phi\right)} & \frac{\delta F\left(\phi\right)}{\delta p\left(\phi\right)}\end{array}\right)\mathcal{D}_{1}\left(\begin{array}{c}
\frac{\delta G\left(\phi'\right)}{\delta r\left(\phi\right)}\\ 
\frac{\delta G\left(\phi'\right)}{\delta p\left(\phi\right)}
\end{array}\right) \ .
\end{equation}
In this way, the brackets of the dynamical fields $r\left(t,\phi\right)$
and $p\left(t,\phi\right)$ at fixed time are given by
\begin{align} \label{field-algebra}
\left\{ r\left(\phi\right),r\left(\bar{\phi}\right)\right\}  & =-\frac{2}{\ell}r\left(\phi\right)\partial_{\phi}^{-1}\left(r\left(\phi\right)\delta\left(\bar{\phi}-\phi\right)\right) \ , \\
\left\{ r\left(\phi\right),p\left(\bar{\phi}\right)\right\}  & =-\frac{1}{\ell}\partial_{\phi}\left(\delta\left(\bar{\phi}-\phi\right)\right)-\frac{2}{\ell}r\left(\phi\right)\partial_{\phi}^{-1}\left(p\left(\phi\right)\delta\left(\bar{\phi}-\phi\right)\right) \ ,\\
\left\{ p\left(\phi\right),p\left(\bar{\phi}\right)\right\}  & =-\frac{2}{\ell}p\left(\phi\right)\partial_{\phi}^{-1}\left(p\left(\phi\right)\delta\left(\bar{\phi}-\phi\right)\right) \ .
\end{align}
In the next subsection, the presented bi-Hamiltonian structure of the AKNS systems will be a key property to be employed in the computation of the Dirac gravitational Poisson brackets.
﻿
﻿
\subsubsection{Abelian algebra}
Following the same spirit as the integrable hierarchy AKNS, the gravitational charged counterpart display an abelian algebra. The proof stems from the fact that the charges canonically generate the transformations \cite{Brown:1986ed}
\begin{equation} \label{charge algebra def}
\{Q[\eta],Q[\bar \eta]\}=\delta_{\bar{\eta}}Q[\eta] \ .
\end{equation} 
We take the variation of our canonical charge \eqref{deltaQ} and consider that the field variations \eqref{eomM} can be written as, 
\begin{equation}
\left(\begin{array}{c}
\delta{r}\\
\delta{p}
\end{array}\right)=\mathcal{D}_{1}\left(\begin{array}{c}
\mathcal{R}_{M+1}\\
\mathcal{P}_{M+1}
\end{array}\right)=\mathcal{D}_{2}\left(\begin{array}{c}
\mathcal{R}_{M+2}\\
\mathcal{P}_{M+2}
\end{array}\right) \ , 
\end{equation}
so that the right hand side of the charge algebra \eqref{charge algebra def} is given by
\begin{equation} \label{charge proof 1}
\{Q[\eta],Q[\bar \eta]\}=\frac{k\ell}{2\pi}  \sum_{n=0}^M \lambda^{M-n}\int d\phi \  
\mathcal{S}_{n+1}^T\  \mathcal{D}_1\  \mathcal{S}_{\bar M +1} \ .
\end{equation}
Here we used that the fact that the two copies are independent and therefore we can focus in a generic case without specifying the upper index introducing the following notation, 
\begin{equation}
\lambda=\lambda^\pm\ , \quad 
 \mathcal{S}_\ell=\begin{pmatrix}
  \mathcal{R}_{\ell}^\pm \\ \mathcal{P}_{\ell}^\pm 
\end{pmatrix}\ , \quad  \mathcal{S}_\ell^T=\begin{pmatrix}
  \mathcal{R}_{\ell}^\pm & \mathcal{P}_{\ell}^\pm
\end{pmatrix} \
\end{equation}
 Because the Poisson brackets are antisymmetric we can write \eqref{charge proof 1} as
\begin{equation}
\{Q[\eta],Q[\bar \eta]\}=-\frac{k\ell}{2\pi}  \sum_{n=0}^M \lambda^{M-n}\int d\phi\  
\mathcal{S}_{\bar M+1}^T\  \mathcal{D}_1\  \mathcal{S}_{ n +1} \ .
\end{equation}
From the recurrence relation \eqref{bihamrec} we can impose the relation $\mathcal{D}_1 \mathcal{S}_{n +1}=\mathcal{D}_2 \mathcal{S}_{n +2}$ so that, 
﻿
﻿
\begin{align}\notag
\{Q[\eta],Q[\bar \eta]\}&=-\frac{k\ell}{2\pi}  \sum_{n=0}^M \lambda^{M-n}\int d\phi \  \mathcal{S}_{\bar M+1}^T\  \mathcal{D}_2\  \mathcal{S}_{ n +2}= \frac{k\ell}{2\pi}  \sum_{n=0}^M \lambda^{M-n}\int d\phi \ \mathcal{S}_{ n+2}^T\  \mathcal{D}_2\  \mathcal{S}_{\bar M +1} \\
&= \frac{k\ell}{2\pi}  \sum_{n=0}^M \lambda^{M-n}\int d\phi \ \mathcal{S}_{ n+2}^T\  \mathcal{D}_1\  \mathcal{S}_{\bar M} \ .
\end{align}
The consecutive application of this procedure $s$ times can be expressed as,
\begin{equation}
 \sum_{n=0}^M \lambda^{M-n}\int d\phi
\  
\mathcal{S}_{n+1}^T\  \mathcal{D}_1\  \mathcal{S}_{\bar M +1}
= \sum_{n=0}^M \lambda^{M-n}\int d\phi
\  
\mathcal{S}_{n+1+s}^T\  \mathcal{D}_1\  \mathcal{S}_{\bar M +1-s} \ .
\end{equation}
Since $s$ is an arbitrary number, one can choose it such that  
so that $n+s=\bar M$,  wich implies $s=\bar M -n$ and we complete the proof that the algebra generated by the charges is abelian,
\begin{equation} 
\{Q[\eta],Q[\bar \eta]\}=\frac{k\ell}{2\pi}  \sum_{n=0}^M \lambda^{M-n}\int d\phi
\  
\mathcal{S}_{\bar M+1}^T\  \mathcal{D}_1\  \mathcal{S}_{n +1}=\{Q[\bar\eta],Q[\eta]\}
\end{equation}
Hence, the generators of the gravitational side inherit the integrable property of the AKNS system; the involution of an infinite tower of charges spanning an Abelian algebra $$\{Q[\eta],Q[\bar\eta]\}=0.$$
﻿
﻿
﻿
﻿
﻿
﻿
﻿
﻿
﻿
﻿
﻿
﻿
﻿
﻿
﻿
﻿
﻿
﻿
﻿
﻿
﻿
﻿
﻿
﻿
﻿
﻿
﻿
﻿
﻿
﻿
\section{Black hole Thermodynamics}
In this section we study black hole solutions applying the Euclidean approach to the general case of the AKNS hierarchy.  We define a geometric temperature fixed by the regularity of solutions and show that is always constant, provided the properties of the spectral polynomial connected to hyperelliptic curves.  We put special care on the KdV equation,  finding a periodic cnoidal wave solution which defines the black hole dynamics.  We compute both the mass and its temperature, as well as its entropy.
﻿
\subsection{Regularity conditions and temperature} \label{Regularity conditions}
﻿
To study black holes in the Euclidean approach, we consider the time coordinate to be Wick rotated $t=-i\tau$, for $0\leq \tau<\beta$. Here
 $\beta$ is the period of Euclidean time and the inverse of the Hawking temperature $\beta=\tfrac{1}{T}$.  It can be computed by demanding regular gravitational configurations,  that reduce to the condition
\begin{equation}\label{holonomia}
\mathcal{H}(A)=\mathcal{P}\exp \left( \int^{\beta}_{0} d\tau \mathcal{A}_{\tau} \right)=\mathcal{P}\exp\left(\beta \mathcal{A}_{\tau} \right)=-\mathbb{I} \ ,
\end{equation}
where $\mathcal{A}$ is a stationary Euclidean connection and $\mathcal{H}(\mathcal{A})$ is an element of the group which becomes trivial under contractible cycles. The details of the Euclidean continuation on the gauge connections are given in the Appendix \ref{Apendice D}. One can show that the condition \eqref{holonomia}, reduce to trivial holonomies for each copy $\exp\left(\beta^{\pm} a^{\pm}_{\tau} \right)=-\mathbb{I}_{2x2}$ where $\beta^{\pm}$ are the left and right temperatures.  The computation simplifies considering gauge transformations that diagonalize the auxiliary gauge relations $a^{\pm}$,  leading to an expression proportional to the diagonal matrix $L_0$,\footnote{Note that the gauge connections $a^{\pm}$ are related to $\mathcal{A}^{\pm}$ by the radial gauge transformation. Then the matrix diagonalization leads to a holonomy that does not depend on the radial coordinate, and the regularization condition is imposed only on the fields and not on the coordinates.},  
\begin{equation}\label{diagonaltransformation}
\mathfrak{a}^{\pm}=g_{\pm}^{-1}a^{\pm}g_{\pm}+g_{\pm}^{-1}dg_{\pm}=(\mp\mathcal{J}^{\pm}d\phi-\tfrac{\eta^\pm}{\ell} dt)L_{0}\ .
\end{equation}
Here $\mathcal{J}^{\pm}=\mathcal{J}^{\pm}(t,\phi)$ and $\eta^\pm=\eta^\pm(t,\phi)$ are functions accompanying $d\phi$ and $dt$ respectively, while $g_{\pm}$ are the corresponding group elements  which can be written as,
\begin{equation}\label{group-element}
g_{\pm}=e^{f^{\pm}L_{\pm}}e^{h^{\pm} L_{\mp}} \ , 
\end{equation}
where $f$ and $ h$ are periodic functions of the angle,  and $g_{\pm}$ are found to be $L_0$ independent as a direct condition of the equation \eqref{diagonaltransformation}.  The transformation property of the holonomy $\mathcal{H}(\mathfrak{a})=g^{-1}(0)\mathcal{H}(a)g(\beta)$ for $g(\tau)=g(\tau+\beta)$ on the relation \eqref{holonomia} and the expliclit use of \eqref{diagonaltransformation} determine the conditions
\begin{equation}
C^{\pm}(f^{\pm})^{2}-2A^{\pm}f^{\pm}-B^{\pm}=0 \ , \qquad h^{\pm}=-\frac{C^{\pm}}{2\sqrt{(A^{+})^2+B^{+}C^{+}}} \ .
\end{equation}
In this way,  we find that the functions $\eta^{\pm}$ have always a constant value, which is a direct consequence of them being proportional to the square root of the spectral polynomial \eqref{spec2},
\begin{equation}
\eta^{\pm} = 2 \sqrt{(A^{\pm})^2+B^{\pm}C^{\pm}} \ .
\end{equation}
Thus, the direct diagonalization of the matrix and the holonomy conditions fix the value of the left and right temperatures as,
\begin{equation} \label{Temperatureforhwgauge}
\beta^{\pm}=\frac{n \pi  \ell }{\sqrt{(A^{\pm})^{2}+B^{\pm}C^{\pm}}}=\frac{2 n \pi\ell }{\eta^{\pm}} \ ,
\end{equation} 
where $n$ an odd integer. The case of the Hawking temperature is given for $n=1$, where $\beta=\tfrac{1}{2}(\beta^{+}+\beta^{-})$. We emphasize that despite the dependence of the metric on the angular variable is quite non-trivial to find that \textit{the temperature is always constant for any integrable stationary AKNS hierarchy} and is given in terms of the spectral polynomial. ﻿Let us  see these results in a more concrete way, using an example based on the KdV stationary hierarchy.
﻿
﻿
\subsection{The KdV black holes and cnoidal gravitational solutions}\label{KDV black hole}
The KdV equation is perhaps the most well-known prototype of a nonlinear integrable equation. The applications of this equation wide from its origin in the motion of a fluid profile in shallow water channel to field theory, quantum mechanics and also three dimensional gravity. In \cite{pttt},  it was shown how asymptotic symmetries can fit within this integrable hierarchy.  
Here we revisit the KdV equation (\ref{ksta}), as boundary conditions of AdS$_{3}$ as a special case of the AKNS hierarchy.  In the context the boundary conditions presented in this article, the role of the spectral parameter is fundamental for the construction of stationary black hole solutions. We stress that the same idea can be applied to any equation in the KdV hierarchy including non-linear integrable equations with higher order derivatives. Following the steps of section \ref{secakns} we find that KdV equation in terms of the dynamical fields of the following form,
 \begin{equation}
\mp\ell \dot{p}^{\pm}=p^{\pm\prime\prime\prime}+6p^{\pm \prime} p^{\pm} \  , \quad  r^{\pm}=-1 \  .
\end{equation}
Thus, the explicit form of the functions that appear in the metric \eqref{compac} and \eqref{metricaAKNS} are,
\begin{subequations}   \label{functions for KDVbh}
 \begin{align}
A^{\pm}&=p^{\pm \prime}{-}2\lambda^\pm{-}4 (\lambda^{\pm})^3 \ , \\ \quad B^{\pm}&=p^{\pm \prime\prime}{+}2 (p^{\pm})^2{-}2\lambda^\pm p^{\pm\prime}{+}4(\lambda^{\pm})^2  p^{\pm}\ ,  \quad  \\C^{\pm}&=-2p^{\pm}{-}4(\lambda^\pm) ^2  \ .
 \end{align} 
\end{subequations}
In the following, we will describe time-independent black hole solutions, $\dot p=0$, defining the stationary KdV equation 
 \begin{equation}\label{kdv1}
p^{\pm\prime\prime\prime}+6p^{\pm \prime} p^{\pm}=0 \  .
\end{equation}
Among the different solutions of the KdV equation we are interested in the periodic ones in the angular variable, which leads to the cnoidal soliton wave \cite{drazin}, 
 \begin{equation}\label{cnoidal}
p^\pm(\phi)=\frac{2 n^{\pm2} K^{\pm 2}}{3\pi^2}\left[\nu^\pm+1-3\nu^\pm\  {\rm sn}^2 \left(\frac{ K^\pm}{\pi} n^\pm \phi ,\nu^\pm\right)\right] \ .
\end{equation}
Here ${\rm sn} (z,\nu)$ is the Jacobi elliptic sine function depending on the modular parameter $0<\nu<1$ with real period $T=2K(\nu)$, where $K(\nu)$ is the complete elliptic integral of the first kind. The soliton solution (\ref{cnoidal}) is $2\pi$-periodic $p^\pm(\phi+2\pi)=p^\pm(\phi)$ where we denote $K^{\pm}=K(\nu^\pm)$ and $n^\pm \in \mathbb{Z}$ is a winding type of the parameter which will describe the black hole solutions. Once we establish the periodicity nature of the solution \eqref{cnoidal}, the stationary equation \eqref{kdv1} can be integrated twice,
 \begin{align}\label{kdv2}
p^{\pm\prime\prime}+3p^{\pm 2}+\mu^\pm&=0 \  , \\
\label{kdv3}
(p^{\pm\prime})^2+2p^{\pm 3}+2\mu^\pm p^{\pm}+\delta^\pm&=0 \  ,
\end{align}
where integration constants $\mu^\pm$ and $\delta^\pm$ are given in terms of the solution parameters,
\begin{align}\notag
\mu^\pm&=-\frac{4n^{\pm4}}{3 \pi^4}(\nu^{\pm2}-\nu^\pm+1) K^{\pm 4} , \\
 \delta^\pm&=\frac{16n^{\pm6}}{27 \pi^6}(\nu^\pm+1)(\nu^\pm-2)(2\nu^\pm-1)K^{\pm 6} \  .
\end{align}
These integration constant are closely related to the spectral polynomial of the hiearchy that can be computed using the equations \eqref{functions for KDVbh}, \eqref{kdv2} and \eqref{kdv3},
\begin{align}
(A^{\pm})^2+B^{\pm}C^{\pm}&=16 \lambda^{\pm6}+4\mu^\pm \lambda^{\pm2}-\delta^\pm\\
&=16\left( \lambda^{\pm2}-E_0^\pm\right)\left( \lambda^{\pm2}-E_1^\pm\right)\left( \lambda^{\pm2}-E_2^\pm\right) \ .
\end{align}
It is easy to check  that the six constants $E_0^\pm$, $E_1^\pm$ and $E_2^\pm$  are explicitly given by
 \begin{equation}
E_0^\pm=\frac{n^{\pm2}(2-\nu^\pm)K^{\pm 6} }{3\pi^2}\ , \quad E_1^\pm=\frac{n^{\pm2}(2\nu^\pm-1)K^{\pm 6} }{3\pi^2} \ , \quad E_2^\pm=\frac{n^{\pm2}(1+\nu^\pm)K^{\pm 6} }{3\pi^2}\ ,  
\end{equation}
and satisfy the following relations,
\begin{align}\notag
E_0^\pm+E_1^\pm+E_2^\pm=0\ , \quad 
4(E_0^\pm E_1^\pm+E_1^\pm E_2^\pm+E_2^\pm E_0^\pm)=\mu^\pm \ , \quad 
\frac{1}{16} E_0^\pm E_1^\pm E_2^\pm= \delta^\pm \ .
\end{align}
The constants $E_\alpha^\pm$ are the singlet energies of the band edge states of the Lam\'e problem associated with the cnoidal solution, see for instance \cite{Correa:2008hc, Correa:2008bz}. This can be seen from the linear problem \eqref{linear} for the two copies in terms of $\partial_\phi \Psi^\pm =a_\phi^{\pm}\Psi^\pm $, where $\Psi^\pm$ is an auxiliary spinor field. After decoupling the linear problems, the main equations to solve for the spinor components $\psi^{\pm}$ are two copies of the Lam\'e equations associated with the Schr\"odinger stationary problems, 
\begin{equation} -\psi^{\pm \prime \prime }-p^{\pm}(\phi)\psi^{\pm}=- (\lambda^{\pm})^2\psi^{\pm} \ , 
\end{equation} where the cnoidal solutions \eqref{cnoidal} play the role of minus times the potential.
﻿
﻿
\subsection*{Cnoidal KDV black hole thermodynamics}
Black hole thermodynamics can be studied by considering the euclidean Chern-Simons gravity action $I^{\text{E}}$, that for on-shell configurations reduces to (see e.g. \cite{Bunster:2014mua})
\begin{equation}
I^{\text{E}}_{\text{on-shell}}=\mathcal{B}(\infty)-\mathcal{B}(r_{+}),
\end{equation}
being $\mathcal{B}$ the boundary term that makes the action well-defined. It is evaluated at surface of infinite radius and at the black hole horizon $r_{+}$, where regularity conditions have to be imposed.  In particular $\mathcal{B}=\mathcal{B}^{+}-\mathcal{B}^{-}$ , for
\begin{equation}
\delta \mathcal{B}^{\pm}=  -\frac{k}{2\pi}\int\displaylimits d\tau d \phi \langle a^{\pm}_{\tau}\delta a^{\pm}_{\phi}\rangle .
\end{equation}
At radial infinity, we expect to obtain terms proportional to the conserved charges,  i.e. $\delta \mathcal{B}({\infty})=-\beta\delta M$.  We recall that since the solution is angle-dependent, the angular momentum is not a conserved.  Indeed,  the conditions to have simultaneously mass and angular momentum as conserved quantities for a black hole solution is presented in the subsection \ref{Symmetry transformations}, and they simply lead to the BTZ black hole.  The mass of a cnoidal KDV black hole can be obtained after considering \eqref{functions for KDVbh}, 
\begin{equation}
M=\frac{k}{2\pi}\oint\displaylimits d \phi \left( p^{+2}+p^{-2}+4 \lambda^{+2} p^+ + 4 \lambda^{-2} p^-\right),
\end{equation}
which evaluated in the cnoidal solution, reduce to
\begin{equation}
M=M^++M^- \ , 
\end{equation}
where
\begin{equation}
M^\pm=\frac{8k\, n^{\pm2} }{18 \pi ^4}K^\pm \left[(\nu^{\pm2}{-}\nu^\pm{+}1) n^{\pm2}  K^{\pm3}+6 \pi ^2 \xi ^2( 3E^\pm{+}(\nu^\pm{-}2) \lambda^{\pm2} K^\pm )\right] \ .
\end{equation}
The entropy of the solution is obtained after evaluating the boundary term at the horizon $\delta \mathcal{B}(r_{+})=\delta S$, once the regularity conditions are imposed on the gauge connections which fixes the value of the temperature, as presented in the previous subsection \ref{Regularity conditions}. For the cnoidal KdV black, one find that the left and right temperatures  \eqref{Temperatureforhwgauge} are given by 
\begin{equation}\label{Temperatureforhwgauge}
\beta^{\pm}=\frac{ \pi  \ell }{\sqrt{16 \lambda^{\pm 6}+4\mu^\pm \lambda^{\pm2}-\delta^\pm}}=\frac{ \pi  \ell }{4\sqrt{\left( \lambda^{\pm 2}-E_0^\pm \right)\left( \lambda^{\pm 2}-E_1^\pm\right)\left( \lambda^{\pm 2}-E_2^\pm\right)}},
\end{equation} 
that have the remarkable property to be the associated eigenvalues of the quantum problem.
 
As shown in the previous subsection, the regularity conditions were studied in the so-called diagonal gauge, which is also useful to compute the entropy. We make use of the relation,
\begin{equation}
 \int\displaylimits d\tau d \phi \langle a^{\pm}_{\tau}\delta a^{\pm}_{\phi}\rangle =\int\displaylimits d\tau d \phi \langle \mathfrak{a}^{\pm}_{\tau}\delta \mathfrak{a}^{\pm}_{\phi}\rangle,
\end{equation}
which means that the boundary term is preserved under the gauge transformations \eqref{diagonaltransformation}. Considering this relation, and in terms of the diagonal components, 
\begin{equation}
\delta S= \frac{k}{4\pi \ell}\oint\displaylimits d \phi \left( \beta^{+}  \eta^{+} \delta \mathcal{J}^{+}+\beta^{-} \eta^{-} \delta \mathcal{J}^{-}\right).
\end{equation}
To integrate, we evaluate the left and right temperatures  \eqref{Temperatureforhwgauge}, 
\begin{equation}
S=\frac{k}{2}\oint (\mathcal{J}^{+}+ \mathcal{J}^{-}).
\end{equation}
The general expression of this entropy coincides with the one obtained for near horizon geometries \cite{Afshar:2016wfy,opt}. In our particular case, to relate the diagonal fields with the cnoidal KdV black hole parameters \eqref{cnoidal}, we use the diagonalization condition \eqref{diagonaltransformation}, that imposes  $f^{\pm}=\frac{1+2\lambda^{\pm}\ h^{\pm}+h^{\pm'}}{2h^{\pm}}$ and $\mathcal{J}^{\pm}=\frac{-(1+h^{\pm'})}{h^{\pm}}$ on \eqref{group-element}.  These choices are enough to connect the dynamics fields in both gauges by means of a Miura transformation, 
\begin{equation} \label{ecuacion-offdiagonal}
  p^{\pm}-(\lambda^{\pm})^2=\frac{(\mathcal{J}^{\pm})'}{2}-\frac{(\mathcal{J}^{\pm})^{2}}{4},
\end{equation}
which was already observed in \cite{Afshar:2016wfy, Cardenas:2024hah}. It is also worthwhile to mention that Einstein's equations in this gauge reduce to 
\begin{equation}
\dot{\mathcal{J}^{\pm}}=\pm \partial_{\phi}\eta^{\pm},
\end{equation}
as in the stationary regime, they imply $\partial_{\phi}\eta^{\pm}=0$. Consequently, it shows an alternative argument to the one offered by the constancy of the spectral polynomial,  reassuring that the black hole temperature remains constant. 
﻿

﻿
﻿
﻿
﻿
\section{Discussion}
	In this article we have expanded the results presented in  \cite{Cardenas:2021vwo}, that proposes an infinite set of asymptotic conditions for asymptotically locally AdS$_{3}$ spacetimes,  where its dynamics leads to two copies of the Ablowitz, Kaup, Newell and Segur integrable hierarchy. Through this correspondence, we have revealed a powerful interplay between exact geometries arising from three-dimensional Einstein equations with  $\Lambda<0$ and a wide class of integrable equations, whose properties manifest in various geometric aspects of the gravitational solutions. Indeed, we presented the crucial role of the integrability conditions in the canonical realization of the asymptotic symmetries. Moreover, the AKNS geometries fall in the asymptotic conditions \cite{Grumiller:2016pqb}, that have been connected with induced Weyl symmetries at the boundary \cite{Imbimbo:1999bj}. Thus, we provide a fully constructive framework for generating a different type of holographic content. Gauge parameters are spanned as powers of the spectral parameter allowing an iterative expansion of the Killing equations, splitting the information into field transformations and infinitely many generators.  One finds recursion operators and construct an infinite collection of conservation laws from the gravitational canonical charges. It is an interesting finding to notice that the role of the spectral parameters go beyond their use as ``dull coordinates'' in the fields expansion \eqref{rec_sol}, (\ref{rec_solaa}).  Since these parameters are not fixed by the field equations, one might initially assume they can be arbitrarily set, for instance, to zero.  This is precisely what was done in earlier remarkable works \cite{pttt,Afshar:2016wfy}, where the role of the spectral parameters was largely overlooked. Interestingly, they reemerge in the definition of the black hole temperature \eqref{Temperatureforhwgauge} for non-axisymmetric solutions. The thermodynamic analysis rules out values of $\lambda^{\pm}$ that render the inverse of the temperature singular which, notably,  correspond to  eigenvalues of the associated quantum problem on the integrable side.
   
Another interesting aspect comes from performing the conformal analysis of spatial infinity in the sense of Penrose \cite{Penrose:1962ij,Penrose:1963}. In there, the hypersurface $\rho=\infty$ is associated to a finite line element $d\bar{s}$, conformally related to the physical metric by $ds^{2} =\rho^{2}d\bar{s}$. In the Brown-Henneaux case, $\bar{g}$ is a fixed metric on the cylinder at spatial infinity. Generalizations have been given by constructing an equivalence class of conformally related boundary metrics $\bar{g}=e^{-2\omega}\tilde{g}$, with $\omega$ an independent function of the radial coordinate \cite{Papadimitriou:2005ii,Compere:2008us,Marolf:2012vvz,
Troessaert:2013fma,Alessio:2020ioh,Rooman:2000zi,Rooman:2000ei,Alessio:2017lps,Ciambelli:2019bzz}. We perform the same asymptotic analysis in the AKNS asymptotic metric \eqref{metricaAKNS}, finding a different manner to associate degrees of freedom to the conformal metric.  The analysis  leads to the induced spacetime\footnote{A completely analogous result is obtained by by simply turning off fields $r^{\pm},A^{\pm},C^{\pm}$
and the spectral parameter $\lambda^{2}$ in \eqref{compac}. },
\begin{align}\label{conformal-metric}
d\bar{s}^{2} & =-\frac{B^{+}B^{-}}{\ell^{2}}dt^{2}+\frac{(B^{-}p^{+}-B^{+}p^{-})}{\ell}dtd\phi+p^{+}p^{-}d\phi^{2}\\
 & =\left(-\frac{B^{+}}{\ell}dt+p^{+}d\phi\right)\left(\frac{B^{-}}{\ell}dt+p^{-}d\phi\right),
\end{align}
which explicitly depends on four fields $p^{\pm}$ and $B^{\pm}$, two of them dynamical, as the equation for flat spacetimes $R(\bar{g})=0$ reduces to
\begin{equation}\label{eq-conformal-boundary}
    \pm  \dot{p}^{\pm} =\frac{1}{\ell} B^{\prime \pm} \  .
\end{equation}
The known chiral equation is recovered for $p^\pm=B^\pm$,  nonetheless, a richer integrable structure is allowed in the metric of the cylinder at spatial infinity.  We have,  for example, the sine-Gordon equation under the choices $p^{\pm}=(w')^{2}$ and $B^{\pm}=\cos(w)$ or the Gardner equation,  that contains mKdV,  $B^\pm=3\alpha u^{3}+3u^{2}+u''$ with $\alpha$ an arbitrary real parameter.  Indeed,  any $1+1$ system of integrable differential equations  that exhibits the property $$B^\pm=\frac{\delta \mathcal{H}^{\pm}}{ \delta p^{\pm}},$$ fits into the equation \eqref{eq-conformal-boundary},  being $\mathcal{H}^{\pm}$ the Hamiltonian density of some associated integrable systems and $\partial_{\phi}=\mathcal{D}$ its corresponding symplectic operator \cite{Olver}.  Another interesting feature is noted when we study the 2D space-time in advance and retarded coordinates.  Let us consider
$u=t/\ell+\phi$ and $v=t/\ell-\phi$,  such that the flat metric in two dimensions is given by $d\bar{s}=-dudv$.  Now,  we redefine  the coordinates as $u=u(t,\phi)$ and $v=v(t,\phi)$  as given by the differential relations  $du=-\frac{B^{+}}{\ell}dt+p^{+}d\phi$ and $dv=\frac{B^{-}}{\ell}dt+p^{-}d\phi $.  One can check that the consistency equation of the coordinate transformation to the flat solution  $$d^{2}u=\partial_{a}\partial_{b} u dx^{a}\wedge dx^{b}=0\qquad d^{2}v=\partial_{a}\partial_{b} v dx^{a}\wedge dx^{b}=0\qquad\text{for}\qquad x^{a}(t,\phi)$$
are precisely the evolution equation at the conformal boundary \eqref{eq-conformal-boundary} .
﻿\newline
One aspect to pursue so that to extend this work to other physical contexts,  has to do with studying generalized Gibbs ensembles in black hole physics,  where solutions are endowed with an infinite tower of conserved quantities.  This has already been explored for the KdV symmetries and their relationship between individual microstates and thermodynamic ensembles \cite{Maloney:2018yrz,Dymarsky:2018lhf,Brehm:2019fyy}. 
﻿﻿\newline
Finally,  it would be worthwhile to set up the same scheme presented in this paper in other gauge theories of recent interest for their connection to integrability or black hole physics.  On the one hand,  4D Chern-Simons theories have brought a lot of attention for its relationship with integrable 2-dimensional field theories,  as they have provided new insights on the definition of integrability and their corresponding affine algebras \cite{Costello:2019tri,Lacroix:2021iit}.   On the other,  for the case of BF theories \cite{Fukuyama:1985gg},  which are understood as the gauge theory formulation of Jackiw-Teitelboim gravity \cite{Jackiw:1984je,Teitelboim:1983ux},  they have had great relevance in black hole physics of near horizon extremal black holes \cite{Navarro-Salas:1999zer}.  Examples of that connection to integrability have been given in \cite{Cruz:1999gd,Filippov:1996ye,Navarro:1998hc,Cardenas:2024hah}.
﻿

\section*{﻿Acknowledgemenets}
The authors would like to thank the Instituto de Física at Pontificia Universidad Cat\'olica de Valpara\'iso, Instituto de Sistemas Complejos de Valparaíso and Instituto de Ciencias F\'isicas y Matem\'aticas at Universidad Austral de Chile for their warm hospitality,  where part of this work was completed.  This research has been partially supported by FONDECYT Grants 1211356 and 1231810.

\appendix
﻿
\section{Construction of the asymptotic Killing vectors from the Chern-Simons gauge parameters} \label{Appendix A}
﻿
The asymptotic Killing symmetries can extracted from their relationship with gauge transformations,
\begin{equation} \label{gauge-transf}
\delta \mathcal{A} = d \Lambda + \left[ \mathcal{A},\Lambda \right].
\end{equation}
Indeed, the first thing to consider is that one can assume a particular form of the gauge field $\Lambda=-\eta^{\nu}A_{\nu}$, such that
\begin{align}
\delta_{\Lambda}\mathcal{A}_{\mu} & =\partial_{\mu}\Lambda+\left[\mathcal{A}_{\mu},\Lambda\right]\\
 & =-\text{\ensuremath{\mathcal{L}}}_{\eta}\mathcal{A}_{\mu}+\text{(e.o.m)}.
\end{align}
Then, diffeomorphisms coincide with gauge transformations up to terms proportional to the field equations. We use this fact to construct the asymptotic Killing symmetries for the AKNS case, which come from a combination of both ``plus'' and ``minus'' sectors of the gauge transformations. They can be extracted from the relation between gauge and the metric fields, considering that $\mathcal{A}^{\pm}$ are related to the dreibein $e$ and spin connection $\omega$ via $\mathcal{A}^\pm=\omega \pm e/\ell$, such that 
\begin{equation}\label{gmunu}
  g_{\mu\nu}=\frac{\ell^2}{2}\left\langle \left(\mathcal{A}_{\mu}^{+}-\mathcal{A}_{\mu}^{-}\right),\left(\mathcal{A}_{\nu}^{+}-\mathcal{A}_{\nu}^{-}\right)\right\rangle=2\left\langle e_{\mu},e_{\nu}
  \right\rangle, 
\end{equation}
where $\langle\  , \  \rangle$ is the invariant bilinear form of the gauge group. Then, since the metric is written as the interior product of the dreibein, one can show that its Killing symmetries are the ones associated to the space-time metric, such that,
\begin{equation}\label{eq parameters}
    \Lambda^{+}-\Lambda^{-}=-\eta^{\mu}\left(\mathcal{A}^{+}_{\mu}-\mathcal{A}^{-}_{\mu}\right)=-\frac{2}{\ell}\eta^{\mu}e_{\mu},
\end{equation}
﻿
To obtain the particular results of this article \eqref{asym killing1}, we have to first compute the boundary conditions \eqref{bdab}  with the radial dependence turned on. We take gauge group element to be
\begin{equation}
b^{\pm}\left(\rho\right)=\text{exp}\left(\text{Log}\left(\pm\frac{\rho}{\ell}\right)\right). 
\end{equation}
This choice connects the gauge fields $\mathcal{A}^{\pm}$ with an asymptotically locally AdS${_3}$ metric \eqref{metricaAKNS}. Their explicit form are, 
\begin{equation}\label{Aphitot}
\mathcal{A}_{\phi}^{\pm}=\mp 2\lambda^{\pm}L_{0}-\frac{\rho}{\ell}p^{\pm}L_{\pm 1}+\frac{\ell}{\rho}r^{\pm}L_{\mp 1},\qquad \mathcal{A}_{\rho}^{\pm}=\pm \frac{1}{\rho}L_{0},
\end{equation}
while the temporal component given by
\begin{equation}\label{Attot}
\mathcal{A}_{t}^{\pm}=\frac{1}{\ell}(-2A^{\pm}L_{0}\pm \frac{\rho}{\ell}B^{\pm}L_{\pm 1} \mp \frac{\ell}{\rho}C^{\pm}L_{\mp 1}).
\end{equation}
The gauge parameter should also go under the radial gauge transformation, as their functional form is taken so that to be of the kind $\Lambda^{\pm}=\ell \mathcal{A}^{\pm}_{t}[\alpha,\beta,\gamma]$. Then,
\begin{equation}\label{Lambdatot}
   \Lambda^{\pm}=-2\alpha^{\pm}L_{0}\pm \frac{\rho}{\ell}\beta^{\pm}L_{\pm 1} \mp \frac{\ell}{\rho}\gamma^{\pm}L_{\mp 1}.
\end{equation}
We propose the vector field $ \eta=\{\eta^{t},\eta^{\rho},\eta^{\phi}\}$ and allow its components to be field-dependent. The remaining step is to solve \eqref{eq parameters} considering \eqref{Lambdatot},\eqref{Aphitot} and \eqref{Attot}, which renders the solutions \eqref{killingss}.
﻿
﻿
﻿
 \section{Conventions of the Euclidean continuation}
\label{Apendice D}
The Euclidean continuation of the temporal coordinate changes the isometries to so$(3,1)$, leaving invariant the Hyperbolic three-dimensional space.  In that case,  the Chern-Simons connection is spanned by so$(3,1)$ generators $\tilde{J}$ and $\tilde{P}$, leading to a Lie-valued connection $\mathcal{A}'=e^{a}\tilde{P}_{a}+\omega^{a}\tilde{J}_{a}$ for $e$ and $\omega$ the euclidean dreiben and spin connection respectively.
﻿
Here $\tilde{J}$ and $\tilde{P}$ satisfy the algebra,
\begin{equation}
\left[\tilde{J}_{a},\tilde{J}_{b}\right]=\delta^{cb}\epsilon_{abc}\tilde{J}_{d},\qquad\left[\tilde{J}_{a},\tilde{P}_{b}\right]=\delta^{cb}\epsilon_{abc}\tilde{P}_{d}\qquad\left[\tilde{P}_{a},\tilde{P}_{b}\right]=-\delta^{cb}\epsilon_{abc}\tilde{J}_{d}.
\end{equation}
where $\epsilon_{123}=1$. In a matrix representation,  they are given by $4\times4$ matrices
\begin{equation}
\tilde{P}_{a}=i \left( \begin{array}{cc}
 \tilde{L}_{a}& 0 \\
0 & - \tilde{L} _{a}
\end{array} \right)
\qquad\qquad\qquad
\tilde{J}_{a}= \left( \begin{array}{cc}
 \tilde{L}_{a}& 0 \\
0 &  \tilde{L}_{a}
\end{array} \right)
\end{equation}
where $a,b=1,2,3$ and $ \tilde{L}$ are the SL$(2,\mathbb{C})$ generators. The so$(3,1)$ connection can be written as 
\begin{equation}
\mathcal{A} '= \left( \begin{array}{cc}
\mathcal{A}& 0 \\
0 & \mathcal{A}^{\dagger}
\end{array} \right)
\end{equation}
where $\mathcal{A}=(\omega_{a}+i e_{a})\tilde{L}^{a}$.  In relation to the Lorentzian connections,  one can consider the equation
\begin{equation} \label{relation gauge connection}
\mathcal{A}^{+}=\mathcal{A} \qquad \mathcal{A}^{-}=\mathcal{A} ^{\dagger}.
\end{equation}
provided the change of basis $\tilde{L}_{1}=iL _{0}$, $\tilde{L}_{2}=\tfrac{i}{2}(L _{1}-L _{-1})$ and $\tilde{L}_{3}=\tfrac{1}{2}(L _{1}+L _{-1})$,  taking the matrix form of  SL$(2,\mathbb{R})$ given in \eqref{matrix-rep} and also extending the components of $A^{\pm}$ in \eqref{bdab} to be complex.

﻿
\end{document}